\documentstyle[11pt,moriond,mine,epsfig,multirow]{article}

\def\be{\begin{equation}}
\def\ee{\end{equation}}
\def\bea{\begin{eqnarray}}
\def\eea{\end{eqnarray}}

\begin{document}

\title{HEAVY FLAVOUR PRODUCTION AT HERA}%

\author{ A. LONGHIN }

\myaddress{Dipartimento di Fisica ``G. Galilei''\\ 
via Marzolo 8, I-35131 Padova, Italy\\
e-mail: longhin@pd.infn.it\\
on behalf of \emph{H1} and \emph{ZEUS} collaborations}

\mymaketitle\abstracts{A selection of topics on open heavy quark 
production at \emph{HERA} are reviewed here. Measurements
of charm fragmentation parameters will be presented together with 
developments in the study of $D^*$ di-jet angular distributions. 
Charm production in deep inelastic scattering (\emph{DIS}) is also discussed. 
Finally we deal with recent measurements of $b$ cross sections 
using impact parameters in both \emph{DIS} and photoproduction regimes.} 

\section{Introduction} 

Colliding $ep$ at a center of mass energy of $\sqrt{s}=296-318~\rm{GeV}$ 
\emph{HERA} provides an interesting environment for testing \emph{QCD} 
predictions on heavy quark production. 
The virtual photon emitted by the incoming lepton provides a clean
probe which, interacting with quarks and gluons in the proton,
can initiate hard processes. The scale of the \emph{QCD}
interaction spans over a wide range of values which are under direct
experimental control. This report will concentrate on a selection
of some recent measurements performed by the \emph{H1} and \emph{ZEUS} 
collaborations with a focus in particular on \emph{open} $c$ and $b$ quark 
production in both \emph{deep inelastic scattering} (\emph{DIS}: $Q^2
>1$  $\rm{GeV}^2$ ) and \emph{photoproduction} ($Q^2 \sim 0$ $\rm{GeV}^2$) regimes.

\section{Charm production}\label{subsec:charm}

\subsection{Fragmentation tests}\label{subsec:kin}
The luminosity accumulated in the first phase of \emph{HERA} running (1992-00, 
$\sim 130\ \rm{pb}^{-1}$) allows the study of particular decay channels
which provide measurements of some phenomenological parameters
used to describe charm fragmentation. \emph{ZEUS} recently presented 
measurements of the branching fraction $f(c\to D_{s1}^+)$ 
\footnote{$D_{s1}^\pm(2536)$ is one of the $L=1$ states of the $cs$ system}
~\cite{ds1}, the \emph{strangeness suppression factor}~\cite{gammas} 
$\gamma_s$ (a parameter of Lund string model which rules the 
relative production of \emph{strange} and non strange $D$ mesons) 
and the $P_v = V/(V+PS)$ ratio~\cite{pv} relating the production of the 
vector 
(spin-1) to the pseudo-scalar charmed mesons. The values shown in tab.1
give support to the hypothesis of \emph{charm fragmentation
universality}: the $c$ quark hadronizes in the same way in
$e^+e^-$ and $ep$ interactions.

\begin{figure}
\begin{center}
\begin{minipage}[c]{0.3\linewidth}
\psfig{figure=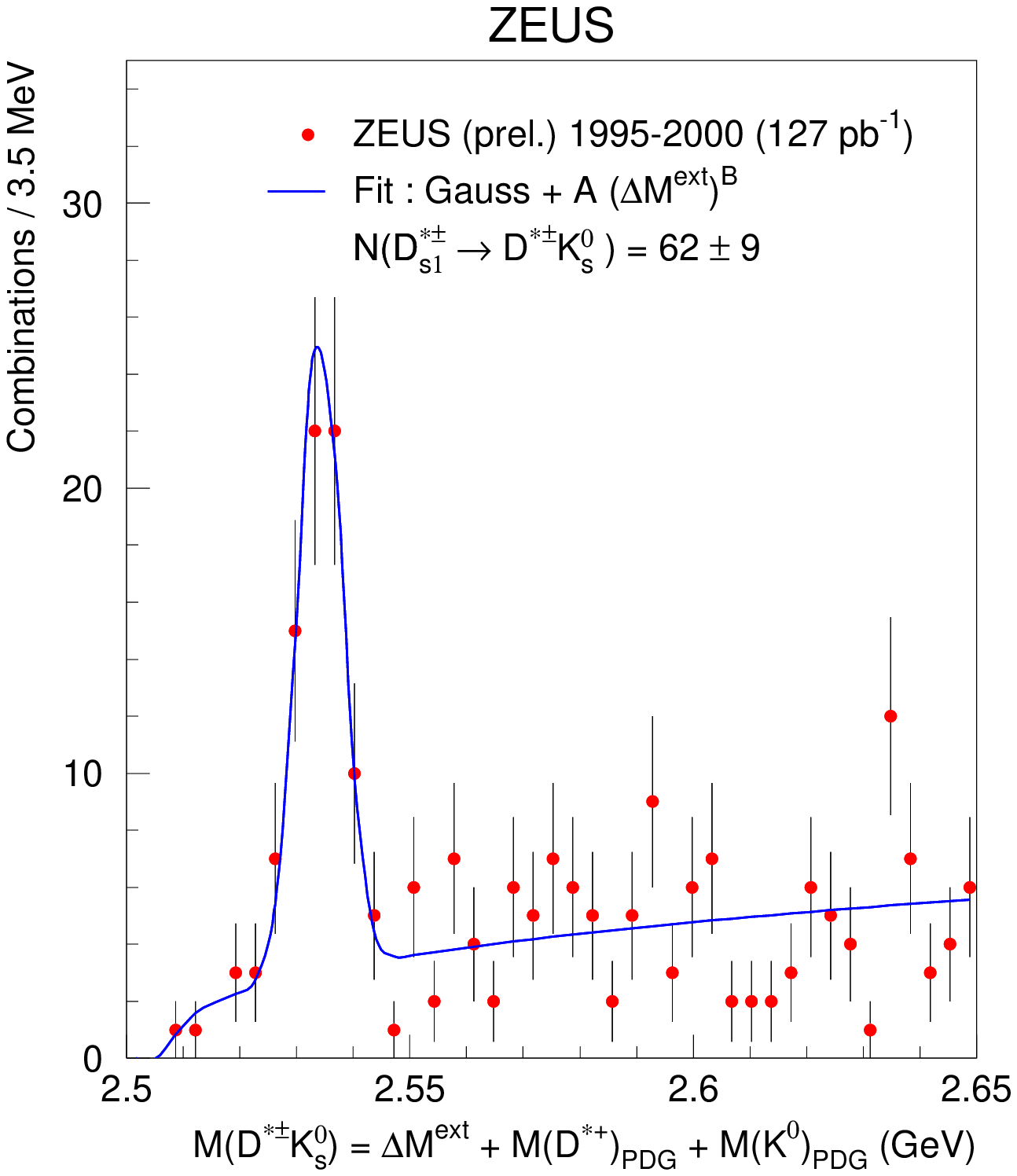,height=2.5in}
\end{minipage}%
\begin{minipage}[c]{0.05\linewidth}
\ 
\end{minipage}%
\begin{minipage}[c]{0.65\linewidth}
\renewcommand{\multirowsetup}{\centering}
\begin{center}
{\footnotesize{Table 1: $c$ fragmentation in $ep$ and $e^+e^-$}}
\vskip 0.5cm
{\scriptsize{
\begin{tabular}{|c|c|c|}
\hline
                                  & $\bf{ep}$                          & $\bf{e^+e^-}$         \\
\hline
\multirow{2}{24mm}{$f(c\to D_{s1}^+)$} & 
\multirow{2}{42mm}{
$(1.24\pm 0.18^{+0.08}_{-0.06}\pm 0.14~(b.r.))~\%$
} &
$(1.6\pm0.4\pm0.3)\ \%$~\cite{lepcfrag} \\ 
 & & $(0.94\pm0.22\pm0.07)\ \%$~\cite{lepcfrag}
\\
\hline
$\frac{\sigma(D_s)}{\sigma(D^*)}$ & $0.41\pm0.07^{+0.02}_{-0.05}$ & $0.43 \pm 0.04$~\cite{gammas}  \\ 
\hline
$\gamma_s = P_s/P_{u,d}$
& $0.27 \pm 0.04^{+0.02}_{-0.03}\pm{0.07}~(b.r.)$               & $ 0.26 \pm 0.03$~\cite{gammas} \\
\hline
\multirow{2}{26mm}{$P_v = D^*/(D^*+D)$} & \multirow{2}{28mm}{$0.546
\pm 0.045 \pm 0.028$} & $0.57 \pm0.05$ ~\cite{pvlep} \\
 & &                    $0.595\pm0.045$~\cite{pvlep}\\
\hline
\end{tabular}
\label{tab:cfrtab}
}}
\end{center}
\vskip 0.5cm
\footnotesize{Figure 1: The $D_{s1}^\pm(2536)$ signal seen in the decay
$D_{s1}^{\pm}\to D^{*\pm}K^0_S$ with $D^{*\pm}\to D^0 \pi^\pm$ and $D^0\to
K^\mp\pi^\pm$}
\label{fig:cfrag}
\end{minipage}
\vskip -0.6cm
\end{center}
\end{figure}

\subsection{Di-jet angular distributions in $D^*$ events}\label{subsec:dijet} 

Di-jet angular distributions, depending on the spin of the exchanged
propagator, are an interesting tool one can use to gain insight into parton
dynamics.  Charm is produced in \emph{direct} processes essentially
through the $q$-exchange diagram $\gamma g\to c\bar{c}$
(\emph{Boson Gluon Fusion}) (fig.2, top right) while \emph{resolved}
production receives contributions also from $g$-exchange processes
like the one shown in the bottom right part of fig.2 ($cg \to
cg$). The distribution of the variable $\cos\theta^* =
\tanh\frac{\eta^{jet1}-\eta^{jet2}}{2}$, $\theta^*$ being the angle
between the beam axis and the di-jet axis in the di-jet rest frame,
has been studied for two different samples enriched in direct or
resolved processes. This separation is defined experimentally by
cutting on the jet-based observable $x_\gamma^{OBS}\equiv
\sum_{jets}{E_T e^{-\eta}/2yE_e}$ which is an estimator of the
fraction of $\gamma$ momentum entering in the hard
scattering. Preliminary results from
\emph{ZEUS}~\cite{ccosthstar}, fig.2,
show a steep angular rise for the resol-
\vskip -0.34cm
\begin{figure}[hpt!]
\begin{center}

\begin{minipage}[c]{0.35\linewidth}
\begin{center}
\psfig{figure=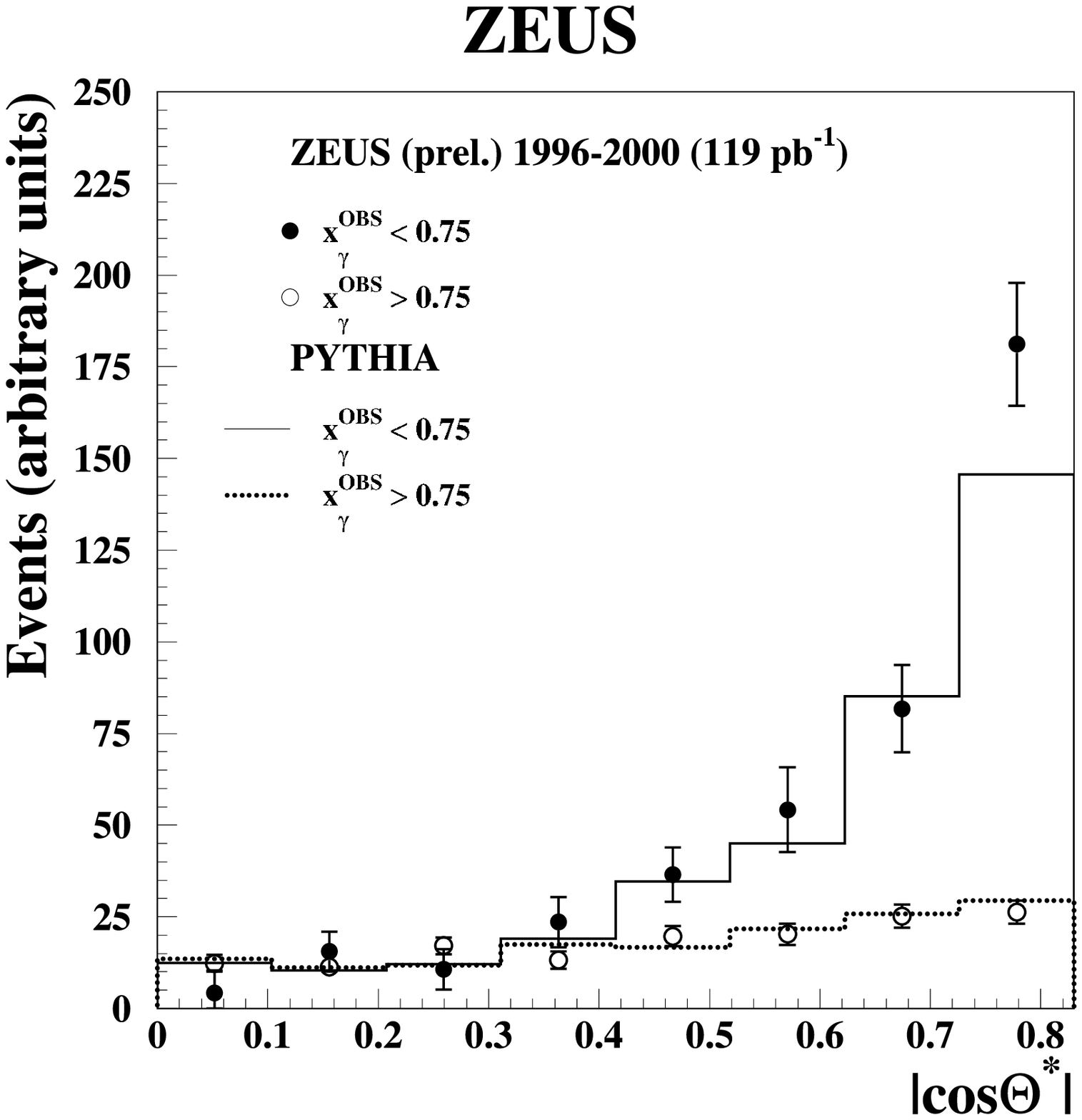,height=2in}
\end{center}
\end{minipage}%
\begin{minipage}[c]{0.15\linewidth}
\begin{center}
\psfig{figure=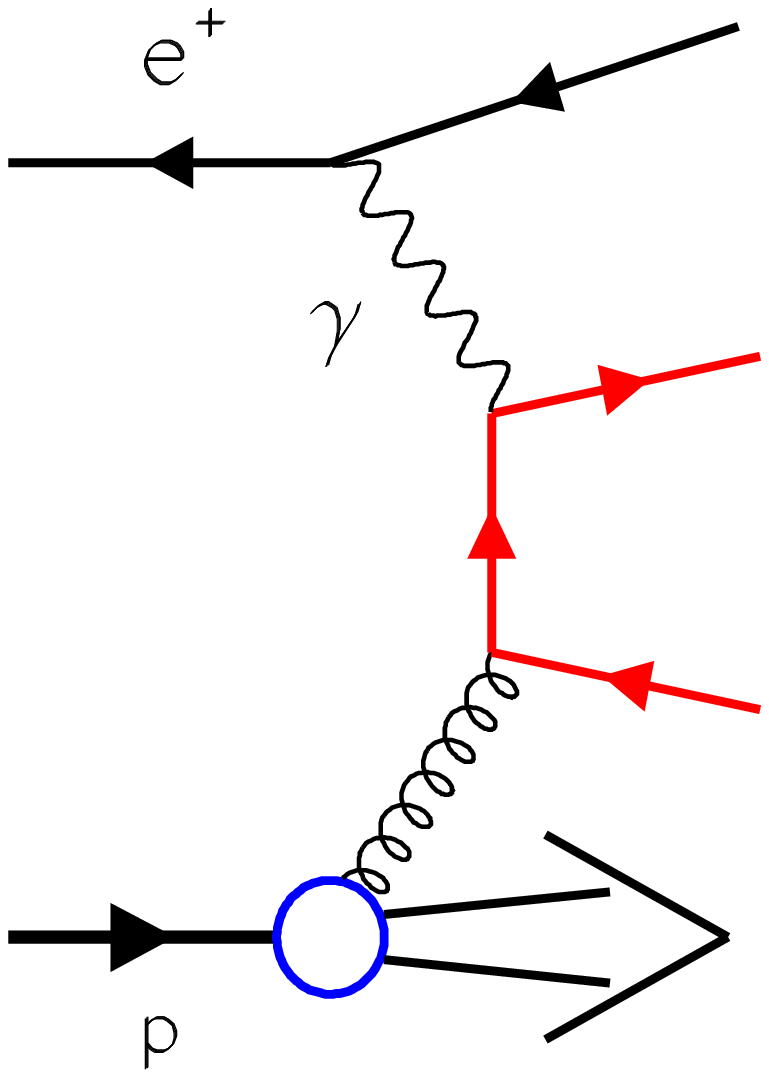,height=0.8in}
\psfig{figure=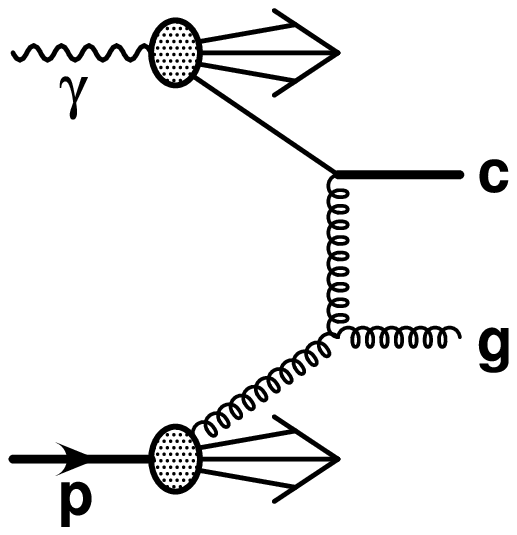,height=0.7in}
\end{center}
\end{minipage}%
\begin{minipage}[c]{0.03\linewidth}
~
\end{minipage}%
\begin{minipage}[c]{0.47\linewidth}
\vskip -0.1cm
ved enriched sample ~ ($x_\gamma^{OBS}<0.75$) 
towards high $\vert\cos\theta^*\vert$ 
in marked contrast to a gentler behaviour 
in the direct enriched sample ($x_\gamma^{OBS}>0.75$). The solid 
histograms are obtained with the \emph{PYTHIA} \emph{LO} Monte Carlo. 
The result is consistent with the fact that the direct processes proceed via
$q$-exchange (spin $1/2$ propagator $\sim
(1-\vert\cos\theta^*\vert)^{-1}$ is expected) while resolved processes 
are dominated by gluon exchange 
(spin $1$ propagator $\sim (1-\vert\cos\theta^*\vert)^{-2}$ 
\emph{Rutherford scattering}). 
This observation is consistent with an impor-
\end{minipage}
\begin{minipage}[c]{0.5\linewidth}
%
\footnotesize{Figure 2: $dN/d\vert \cos\theta^*\vert$ distributions in $D^*$ di-jet 
($E_T>5~\rm{GeV}$) events. The distributions for resolved (black dots) and direct 
(open dots) events have been normalized to each other in the lowest four bins.
\label{fig:dij}}
\end{minipage}%
\begin{minipage}[c]{0.03\linewidth}
~
\end{minipage}%
\begin{minipage}[c]{0.47\linewidth}
tant gluon exchange contribution which is directly associated to the presence of
$c$-excitation processes in the quasi-real photon.
\end{minipage}
\vspace{-0.5cm}
\end{center}
\end{figure}

\subsection{Open charm in DIS: contribution to the $F_2$ structure function}
\label{subsec:F2}

Both \emph{H1} and \emph{ZEUS} measured $F_2^c$, 
the charm contribution to the $F_2$ proton structure function~\cite{f2ch1,f2cz1,f2cz2}:
$\frac{d^2\sigma^{ep \to e c X}}{dxdQ^2} =
\frac{2\pi\alpha^2}{xQ^4}(1+(1-y)^2)\cdot F_2^{c}(x,Q^2)$.
Two procedures for tagging charm have been exploited:
the presence of $D^{*\pm}$ mesons~\cite{f2ch1,f2cz1} or electrons~\cite{f2cz2} 
from $c$ semi-leptonic decays. After the signal has been identified the numbers 
of events in bins of $x$ and $Q^2$ are converted into an inclusive charm cross 
section extrapolating to the full phase space by means of Monte Carlo
generators. Theoretical models are then used to relate the measured cross
section to $F_2^{c}$. The plot in
fig.3a
shows how $F_2^c$ exhibits evident
scaling violations (i.e. $Q^2$ dependence). The ratio of $F_2^c$ to the 
inclusive $F_2$ is presented in fig.3b
as a function of $x$ in $Q^2$ bins. Charm contribution to \emph{DIS}
is definitely sizeable ranging from $\sim 10\ \%$ of the inclusive
$F_2$ at low $Q^2$ up to $\sim\ 40\%$ at $Q^2\sim 500\ \rm{GeV}$ and
$x \sim 0.01$.  This asymptotic contribution at high $Q^2$ is
consistent with the picture in which the $c$ quark can be treated as
any other massless quark ($Q^2 >> m_c$), the relative contribution
following just from a simple charge counting rule. The drop of $F_2^c/F_2$ 
at high $x$ is related to the steep decrease of the proton gluon
density with $x$ leading to a suppression of gluon initiated processes. 
The \emph{HVQDIS}~\cite{refhvqdis} program which
is based on \emph{DGLAP} evolution equations, evaluating the \emph{BGF}
diagram at \emph{NLO}, provides an overall satisfactory description of
$F_2^c$ data.  In addition to \emph{HVQDIS} \emph{H1} uses the
Monte Carlo program \emph{CASCADE}~\cite{casc} which implements the 
so called \emph{CCFM} evolution scheme. Using this approach the
description of data improves at lower values of $x$ and $Q^2$
(not shown).

\begin{figure}[hbpt!]
\begin{center}
\psfig{figure=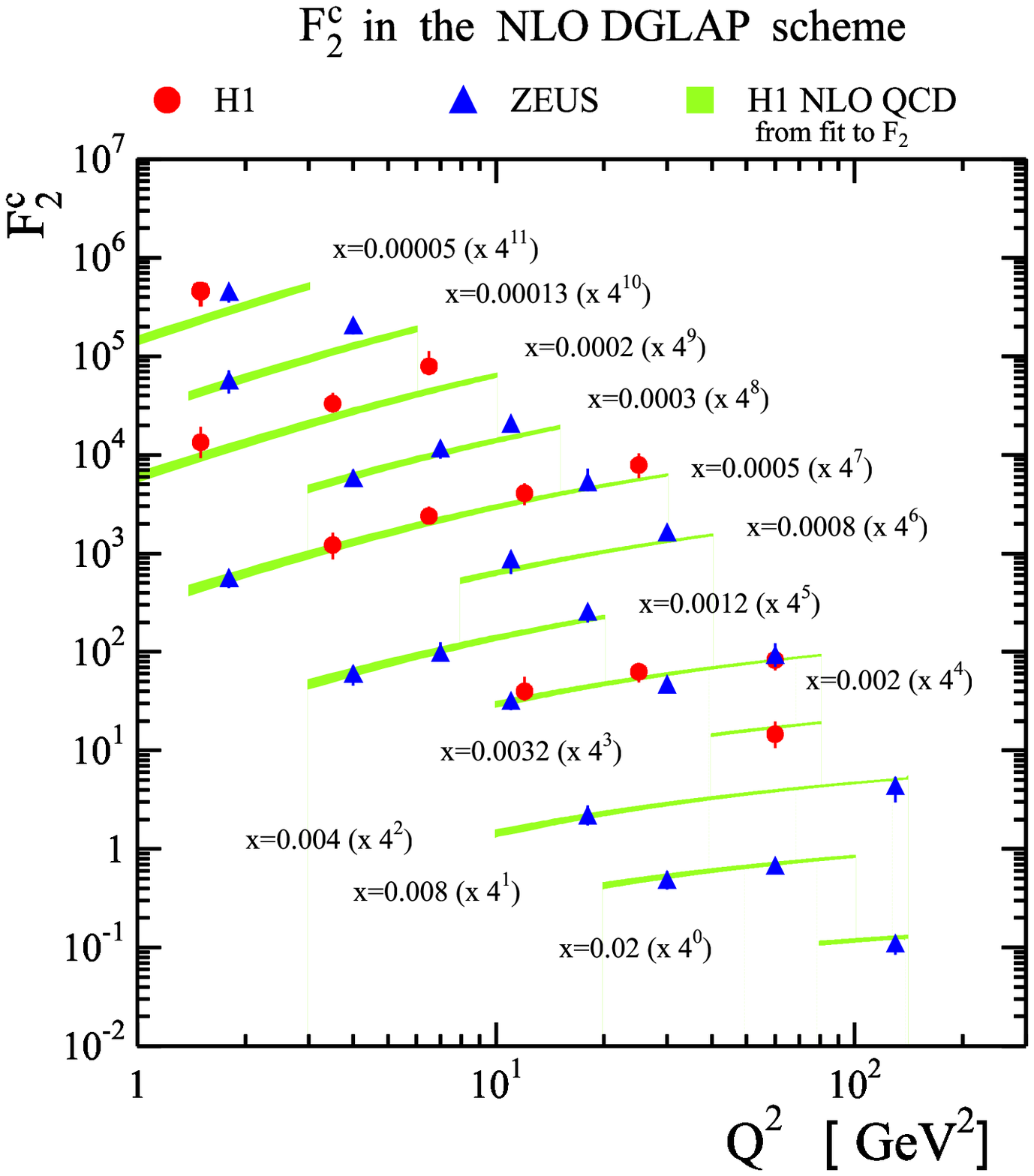,height=3in}
\psfig{figure=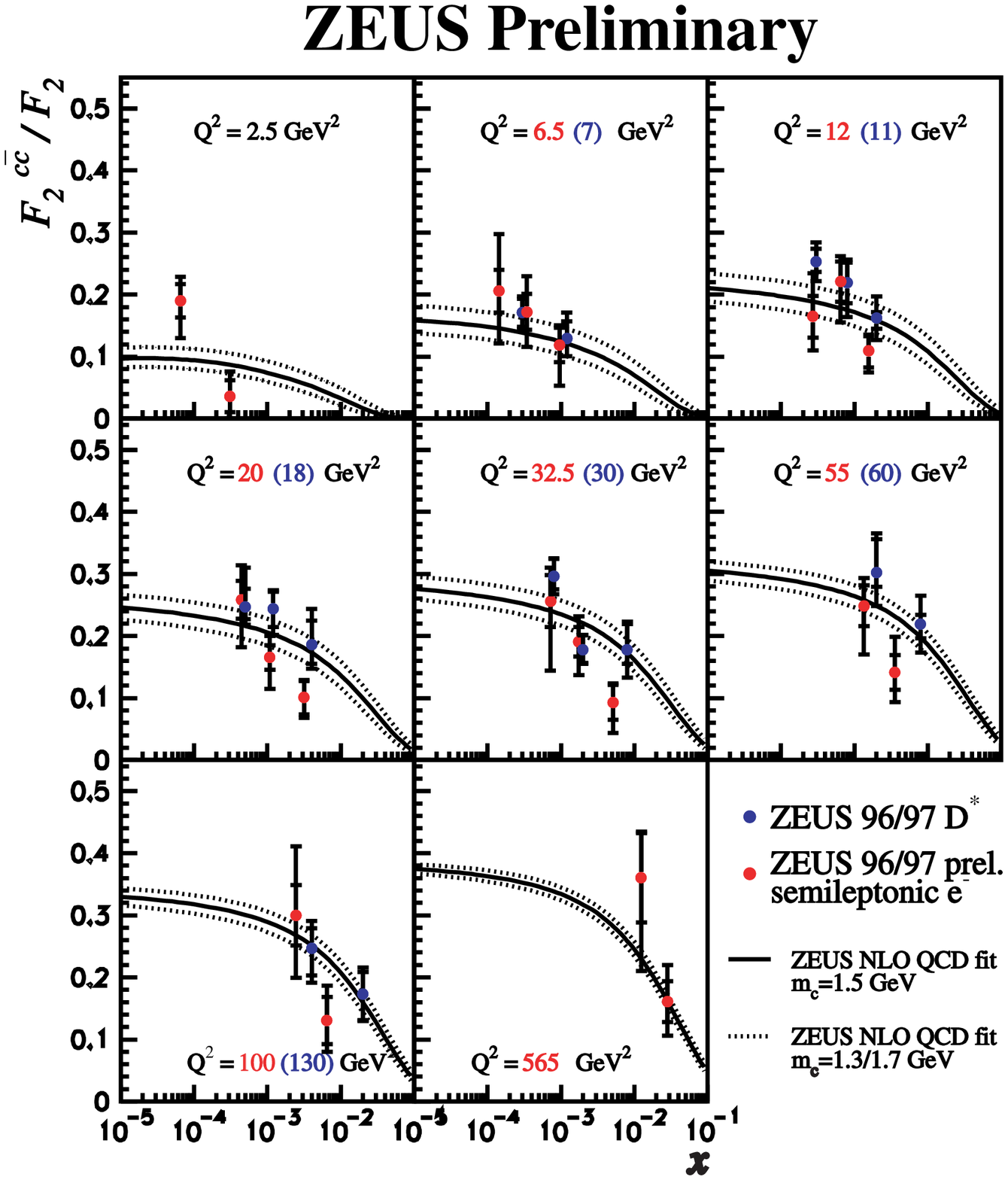,height=3in}\\
\footnotesize{Figure 3: charm contribution to $F_2$ \label{fig:F2c}}
\end{center}
\end{figure}

\section{Beauty production}\label{subsec:beauty}
Both \emph{H1} and \emph{ZEUS} have published results on $b$
production ~\cite{bH1phpold,b1,bH1php,bH1dis,bxdiff}.  The basic
sample used for this measurement consists of events with jets and
moderately high momentum leptons. Due to the high $b$ quark mass, the
lepton from $b$ semi-leptonic decay tends to emerge at higher
transverse momenta with respect to the jet axis ($p_T^{rel}$) than in
the case of $c$ or light quarks decays. This feature allows signal
extraction on a statistical basis by fitting data with Monte Carlo
distributions. The latest \emph{H1} results benefit also from the
presence of a micro-vertex detector information.  The impact parameter
distribution of candidate $\mu$ tracks ($\delta$) is endowed with an
asymmetric tail at positive values (i.e. vertex is downstream of the
associated jet) coming from the presence of long living
particles. This independent signature provides results which are in
good agreement with those obtained using the $p_T^{rel}$ method. The
distributions of the two observables $\delta$ and $p_T^{rel}$ for a
\emph{DIS} selection ($2<Q^2<100~\rm{GeV}^2$, $0.05<y<0.7$,
$p_T^\mu>2\ \rm{GeV/c}$, $30^\circ<\theta^\mu<135^\circ$) are shown in
fig.4.
The measured cross section ($39 \pm 8 \pm 10$) pb \footnote{In the
following first quoted error is statistical and the second
systematic.} lies significantly above the value of \emph{HVQDIS}
\emph{NLO} calculation ($11\pm 2$) pb. The \emph{LO} Monte Carlo 
\emph{AROMA} gives a prediction of $9$ pb and \emph{CASCADE} expects $15$ pb.  The \emph{NLO} theoretical error has been evaluated by varying the
renormalization and factorization scales, $m_b$ and fragmentation
parameters.  Similarly \emph{H1} measured the cross section at low
$Q^2$.  In the following kinematic region: $Q^2<1~\rm{GeV}^2$,
$0.1<y<0.8$, $p_T^\mu>2~\rm{GeV/c}$, $30^\circ<\theta^\mu<135^\circ$
the measured cross section is $\sigma_{vis} = (160\pm16\pm 29)$ pb.
When combined to a previous measurement which used just the
$p_T^{rel}$ variable the result becomes: $\sigma_{vis} = (170\pm 25)$
pb which is well in excess with respect to various expectations which
amount to $38$, $67$, $(54 \pm 9)$ pb for the \emph{AROMA},
\emph{CASCADE} and \emph{NLO} \emph{FMNR} calculations respectively.
The first measurement of $b$ differential cross sections~\cite{bxdiff} has
\begin{figure} \begin{center}
\begin{minipage}[c]{0.65\linewidth}
\begin{flushleft}
\psfig{figure=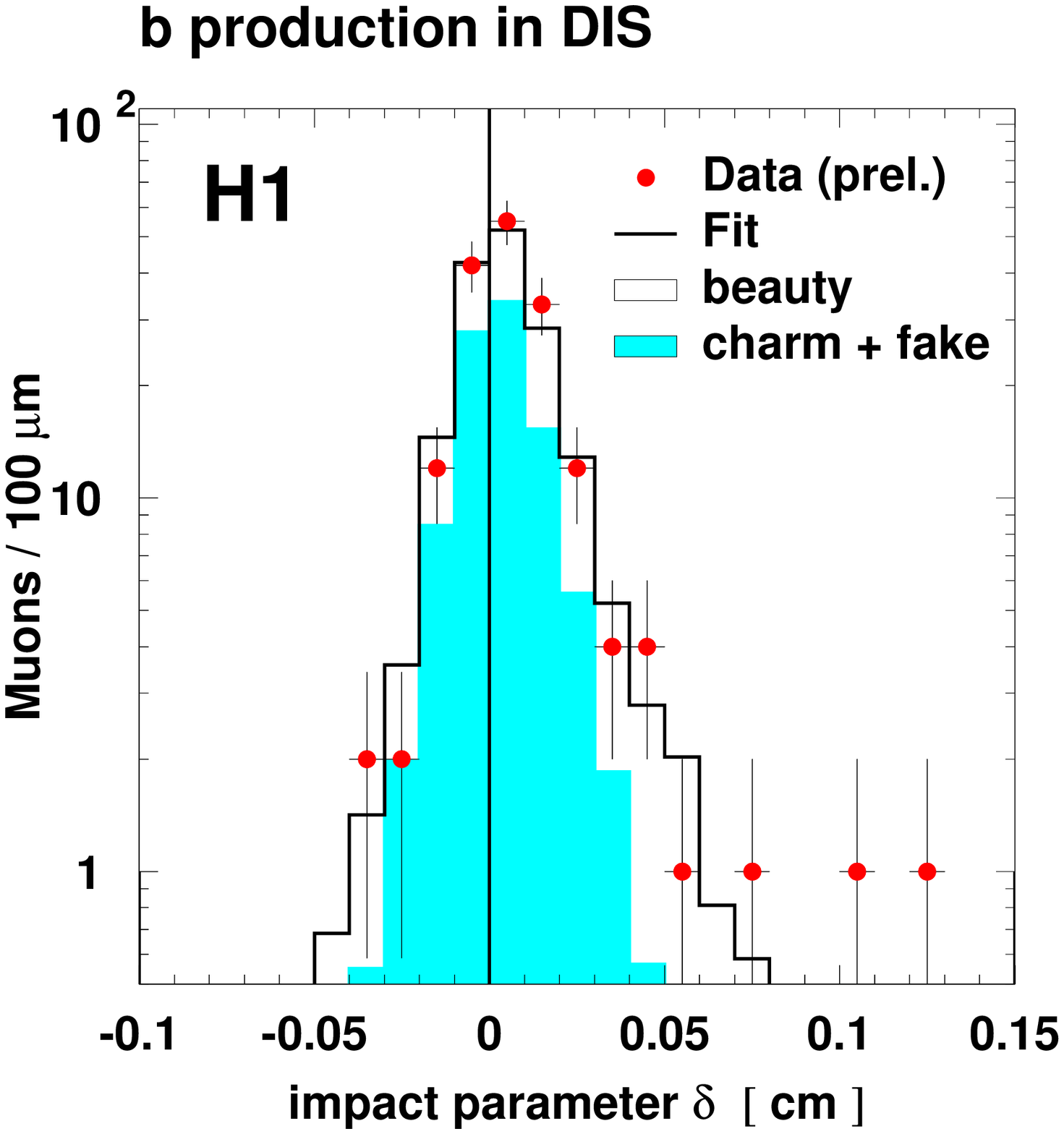,height=2in}
\psfig{figure=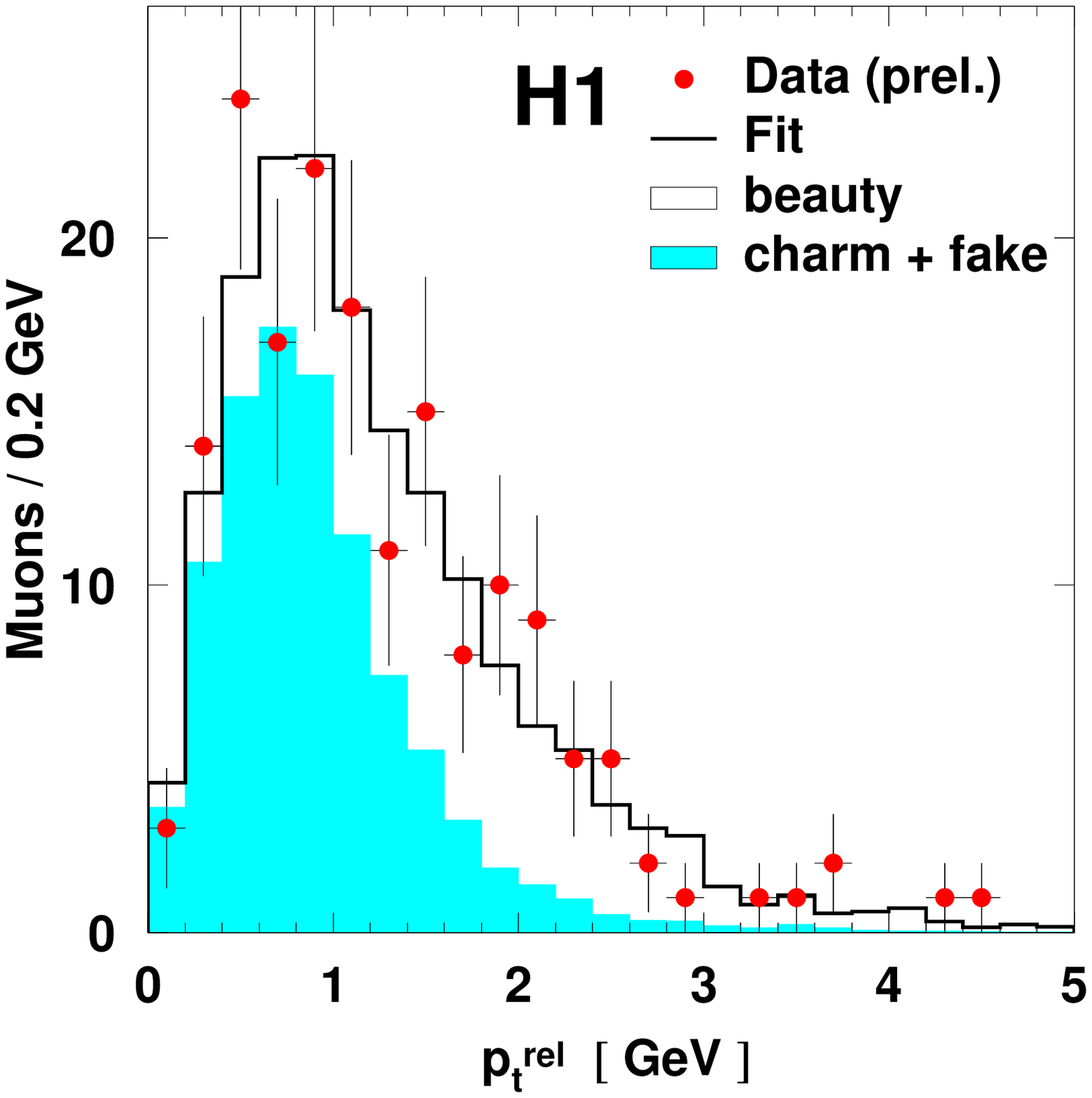,height=2in}
\end{flushleft}
\end{minipage}%
\begin{minipage}[c]{0.35\linewidth}
\footnotesize{Figure 4: $b$ in \emph{DIS}: the solid histogram 
is the fit to data. It is obtained by adding the fitted $b$ signal 
($f_b = (43 \pm 8)\ \%$) to the shaded component which represents the 
$c$ $+$ fake muons background.} 
\label{fig:bDIS}
\end{minipage}%
\vskip -0.6cm
\end{center} \end{figure}

\begin{figure}[hbpt!] \begin{center}

\begin{minipage}[c]{0.67\linewidth}
\vspace{-0.9cm}
also been recently carried out by \emph{ZEUS}. Visible cross sections
in the muon transverse momentum and pseudo-rapidity have been 
calculated. The signal component was determined in each bin through a fit to the 
$p_T^{rel}$ distribution. In this case \emph{PYTHIA} expectation is not far from data 
with some deficit at high $\mu$ pseudo-rapidities where excitation contribution is 
expected to be large.
\emph{HERA} results on $b$ cross sections are summarized in fig.5.
The ratio of the measured cross sections to the predictions at \emph{NLO} is plotted 
for different $Q^2$ regimes. The inner (outer) error bands represent 
the statistical (total) experimental error, the shaded band covers 
the theoretical uncertainty.
\end{minipage}%
\begin{minipage}[c]{0.03\linewidth}~\end{minipage}%
\begin{minipage}[c]{0.3\linewidth}
\vspace{-0.6cm}
\psfig{figure=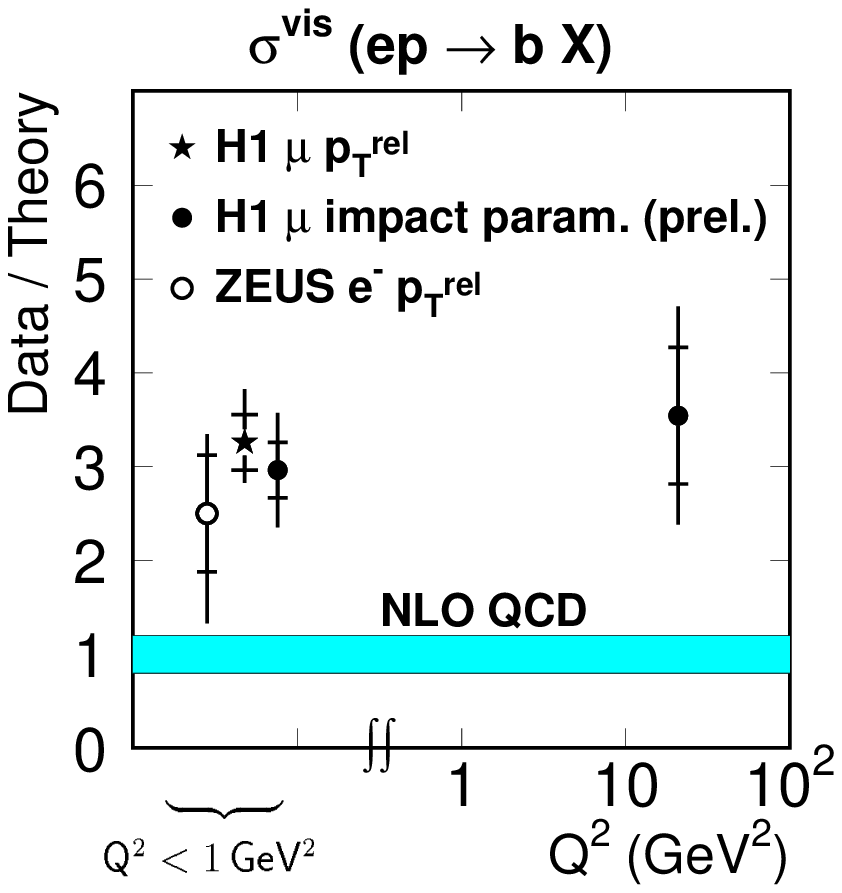,height=2in}
\footnotesize{Figure 5: $b$ results from \emph{HERA}: data/theory.} 
\label{fig:bsum}
\end{minipage}
\vspace{-1.5cm}
\end{center} \end{figure}

\section*{Acknowledgments}
\vskip -0.3cm
I would like to thank my \emph{ZEUS} and \emph{H1} colleagues for the 
suggestions I had from them for the preparation of this talk.

\end{document}